\documentclass[aps,prl,twocolumn,showpacs,preprintnumbers,amsmath,amssymb]{revtex4}

\usepackage{graphicx}
\usepackage{dcolumn}
\usepackage{bm}

\begin{document}

\title{Compaction dynamics of metallic nano-foams: A hydrodynamics simulation study}

\author{J. B. Elliott, M. A. Duchaineau, T. Dittrich, A. V. Hamza, R. Macri and M. Marinak}
\affiliation{Lawrence Livermore National Laboratory,
Livermore, CA 94550, USA\\}

\date{\today}
\begin{abstract}
The compression of a low-density copper foam was simulated with a radiation, hydrodynamics code. In one simulation, the foam had a density of $\rho_0 = 1.3407$ g/cm$^3$, 15\% the density of copper at standard temperature and pressure, and was composed of a tangle of standard density cylindrical copper filaments with a diameter of $4.0\times10^{-7}$ cm. In another simulation, the foam was a uniform material at the same density.  The propagation velocity of the shock wave ($U_f$) through the foam was measured and compared with experimental results.  The simulations show approximately the same agreement with experimental results for $U_f$ and agreed with estimates from different equations of state and simulations using molecular dynamics.  Behavior of the foam ahead of the shock wave is also discussed where the porous nature of the foam allows for the formation of a vapor precursor.
\end{abstract}

\preprint{LLNL-JRNL-468972-DRAFT}
\maketitle

\section{Introduction}

There is interest in using low-density metallic foams at the National Ignition Facility (NIF), but little detailed understanding of the behavior of such foams at conditions accessible at NIF.  It is unknown whether current hydrodynamic simulation codes and equations of state (EOS) can accurately describe the behavior of such foams. 

This work explores the behavior of low-density copper foams in regimes of pressure, density and temperature ($p$, $\rho$, $T$) accessible at the NIF using simulations performed with HYDRA \cite{marinak-01}.  HYDRA is a single fluid, multi-block, multi-material ALE radiation hydrodynamics code.  The radiation transport was treated in the diffusion approximation using a single energy group.  Opacities were calculated from the LEOS tables. The equations of state (EOS) for all materials used were from a combined analytic / Thomas-Fermi EOS which uses a modified Cowan model for the ion EOS, and uses a scaled Thomas-Fermi table for the electron EOS \cite{more-88}.

The results from the HYDRA simulations are compared with experimental data, the results from molecular dynamics simulations of the same system \cite{duchaineau-10} and calculations based on an ideal vapor equation of state and a Van der Waals equation of state.

\section{Simulation details}

\begin{figure}[ht]
  \includegraphics[width=8.7cm]{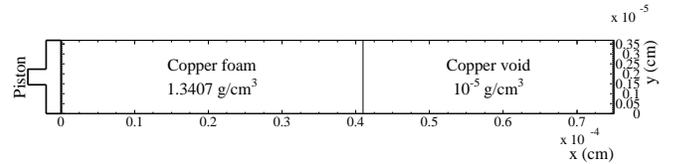}
  \caption{A schematic of the set up of the simulation is shown above.  A piston, moving from left to right, compresses the low-density copper foam.  Downstream of the copper foam is uniformly distributed copper at a density of $10^{-5}$g/cm$^3$ to approximate a vacuum.}
\label{unifoam-schem}
\end{figure}

The system studied was a slab of copper foam with a density of $\rho_0 = 1.3407$ g/cm$^3$ (or 15\% of the normal density of copper $\rho_{\rm bulk} = 8.938$ g/cm$^3$) followed by a {\it copper void} as shown in Fig.~\ref{unifoam-schem}.

The foam slab had the dimensions $36.8\times10^{-7}$ cm $\times~36.8\times10^{-7}$ cm $\times~410\times10^{-7}$ cm.  The foam was compressed by a piston using a constant velocity source in HYDRA that was incident from $x = 0$ and moving in the $+x$ direction.  Piston velocities of $U_p$ = 0.2, 2.0. 20.0 and 50.0 km/s were used in the simulations. The {\it copper void} was copper with a density of $10^{-5}$ g/cm$^3$ and was used to approximate a vacuum in HYDRA.

\begin{figure}[ht]
  \includegraphics[width=8.7cm]{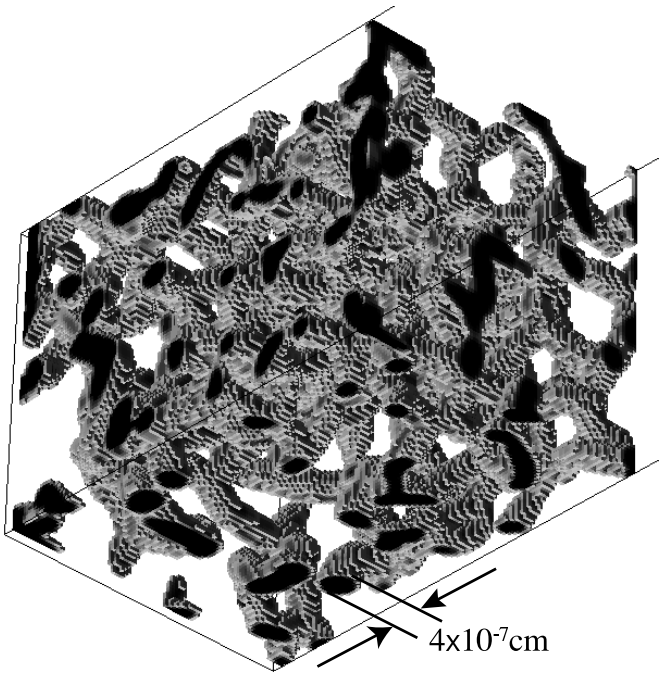}
  \caption{A section of the copper foam.}
\label{filaments}
\end{figure}

Two sets of HYDRA simulations were performed.  One set was three-dimensional and the copper foam was composed of a tangle of standard density copper filaments of cylindrical cross section with a diameter of $4.0\times10^{-7}$ cm; approximately 14 copper atoms reached across a filamentÕs diameter (the radius of a copper atom is $r_{\rm Cu} \approx 1.45 \times 10^{-8}$ cm).  There were approximately ten zones across the diameter of a filament, thus there were approximately 1.4 atoms in each zone of a filament.\ \ Between the filaments was the copper void.\ \ Figure~\ref{filaments} shows a section of the low-density copper foam.  This kind of copper foam was first studied using molecular dynamics \cite{duchaineau-10}.

In the directions transverse to the velocity of the piston, there were 92 zones with lengths of $4.0\times10^{-8}$ cm.  In the foam region, the length of a zone in the longitudinal direction of the piston velocity was also $4.0\times10^{-8}$ cm.  In the copper void region the zone lengths were larger in the longitudinal direction, but the same in the transverse directions.  The simulations had approximately ten million zones.

The second set of HYDRA simulations were performed in two dimensional $RZ$ geometry.  The copper foam was simulated as a material that was completely uniform at density $\rho_0$.  The zoning in the direction along the velocity of the piston was the same as in the three-dimensional simulations.

The simulations were performed in parallel with 64 to 256 processors depending on the number of zones in the simulation.  Simulations with a piston velocity of $U_p = 20.0$ km/s and foam slab lengths of $l_{\rm foam} = 41.0\times10^{-6}$ cm are discussed in detail and results for other values of $U_p$ and $l_{\rm foam}$ reported.

\section{Simulation results}

\subsection{Front velocity}

The main goal of this work was to determine the front velocity $U_f$ that resulted from the compression of the foam by a piston moving at a given velocity $U_p$ in order to test the understanding of the equation of state (EOS) of low-density copper foam and compare with experimental results.

\begin{figure}[ht]
  \includegraphics[width=8.7cm]{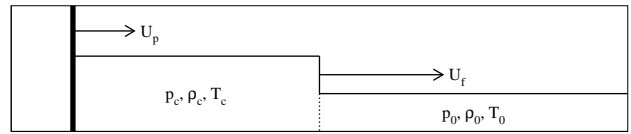}
  \caption{A schematic of a uniform material with initial conditions $p_0$, $\rho_0$, $T_0$ compressed by a piston (thick vertical line) moving with velocity $U_p$.  The compression produces a shock front moving at velocity $U_f$ and compresses material to the state $p_c$, $\rho_c$, $T_c$.}
\label{hug-schem}
\end{figure}

Figure~\ref{hug-schem} shows, schematically, the compression of a uniform material and the resulting shock front.  Given the velocity of the piston $U_p$, the initial pressure, density and temperature ($p_0$, $\rho_0$, $T_0$) of the material and some knowledge of the EOS of the material, one can use the so-called jump conditions to determine the velocity of the shock front $U_f$ and the pressure, density and temperature ($p_c$, $\rho_c$, $T_c$) of the compressed material \cite{zeldovich-02}.

The jump conditions arise from consideration of the conservation of mass:
\begin{equation}
\rho_c \left( U_f - U_p \right) = \rho_0 U_f
\label{mass}
\end{equation}
conservation of momentum:
\begin{equation}
\rho_0 U_f  U_p = p_c - p_0
\label{momentum}
\end{equation}
and the conservation of energy:
\begin{equation}
\rho_0 U_f  \left( \varepsilon_c - \varepsilon_0 + \frac{U_f^2}{2} \right) = p_c U_p .
\label{energy}
\end{equation}
Here $\varepsilon$ is the specific internal energy.

An experiment that measures $U_f$ for a given $U_p$ for a material where the initial pressure, density and temperature values are know then provides some test of the knowledge of the EOS of the material.  To that end, both experiments and simulations commonly measure the front velocity as a function of piston velocity.

For the simulations here, the initial density of the foam was $\rho_0 = 1.3407$ g/cm$^3$ and the initial temperature was $T_0 = 2.56915\times10^{-5}$ keV (room temperature).

The front velocity was determined by dividing the length of the foam by the time at which the front {\it breaks out} of the foam region and into the vacuum region $t_{\rm BO}$
\begin{equation}
U_f = \frac{l_{\rm foam}}{t_{\rm BO}} .
\label{botime}
\end{equation}
The time at which the front {\it breaks in} to the foam region is $t = 0$.\ \ The break out time was defined as the time at which the density in first zone in the vacuum region was equal to the uncompressed foam density.

\begin{figure}[ht]
  \includegraphics[width=8.7cm]{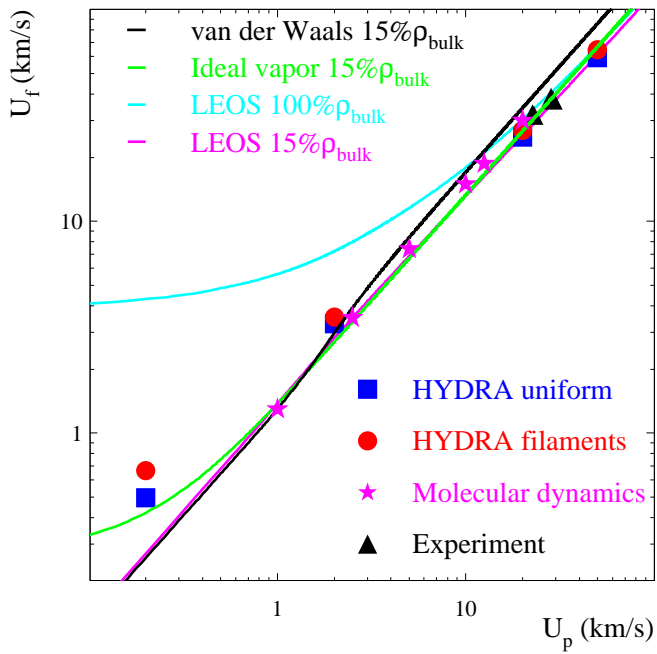}
  \caption{The front velocity is shown as a function of piston velocity: $U_f \left( U_p \right)$.  Solid black triangles show experimental results.  Solid red circles show HYDRA filament simulation results.  Solid blue squares show the HYDRA uniform simulation results.  Solid magenta stars show the results form molecular dynamics simulations. The violet curve shows the LEOS table data for 15\% normal density copper.  The cyan curve shows the LEOS table data for normal density copper.  The green, dashed curve shows the ideal vapor approximation for copper at 15\% normal density. The green curve shows the results using an ideal vapor EOS. The black curve shows the results using a Van der Waals EOS.}
\label{d3-vfvp-pro}
\end{figure}

Values of $U_p =$ 0.2, 2.0, 20.0 and 50.0 km/s were used in the simulations.  The values of $U_f$ for those simulations are shown in Fig.~\ref{d3-vfvp-pro} and listed in Table~\ref{uf-up-res}.  Error bars listed for the front velocity arise from the width of the zones in the simulations and the time intervals at which data from the simulations was recorded.

Figure 4 shows the results of these simulations compared with experimental results, the Hugoniot for copper at 100\% $\rho_{\rm bulk}$ and 15\% $\rho_{\rm bulk}$ from the LEOS tables, molecular dynamics simulation results as well as estimates of $U_f \left( U_p \right)$ using other equations of state and the jump conditions.  For the piston velocities relevant to NIF ($U_f \gtrsim 20$ km/s), the simulations (for both uniform and filament foam) roughly agree with the LEOS table for 15\% $\rho_{bulk}$ and with experimental results \cite{page-09}.

At piston velocities of less than about 10 km/s the difference between copper at 100\% $\rho_{\rm bulk}$ and 15\% $\rho_{\rm bulk}$ from the LEOS tables grows quickly.  Both HYDRA simulations agree more with the 15\% $\rho_{\rm bulk}$ from the LEOS tables than with the 100\% $\rho_{\rm bulk}$ from the LEOS tables.  This is to be expected for the HYDRA uniform foam simulation since that simulation was of uniform copper foam of density 15\% $\rho_{\rm bulk}$ whose behavior was governed by the LEOS tables.

\begin{table}[h]
\begin{center}
\begin{tabular}{|r|c|c|r|r|r|}
\hline
	$U_p$	&	$U_f$			&	$U_f$		&	$U_f$	&	$U_f$	&	$U_f$	\\
	(km/s)	&	(km/s)			&	(km/s)		&	(km/s)	&	(km/s)	&	(km/s)	\\
			&	uniform			&	filaments		&	IV		&	VdW		&	MD		\\
\hline
	0.2		&	$0.496\pm0.003$	&	$0.66\pm0.01$	&	0.42		&	0.26		&	N/A		\\
	2.0		&	$3.28\pm0.001$	&	$3.53\pm0.02$	&	2.69		&	2.97		&	N/A		\\
	20.0		&	$25.09\pm0.08$	&	$26.9\pm0.1$	&	26.67	&	34.29	&	30.0		\\
	50.0		&	$59.5\pm0.4$		&	$64.5\pm0.9$	&	66.67	&	85.84	&	N/A		\\
\hline
\end{tabular}
\end{center}
\caption{Simulation results for the front velocity as a function of piston velocity for a copper foam at $15 \% \rho_0$.}
\label{uf-up-res}
\end{table}

The agreement between the HYDRA filament foam simulation and the 15\% $\rho_{\rm bulk}$ from the LEOS tables is more significant.  It illustrates that the behavior of a more physically realistic foam is well described by the LEOS tables in regimes important for NIF experiments.

The stars in Fig.~\ref{d3-vfvp-pro} show the results from the molecular dynamics simulations for values of the piston velocity of $U_p = 1, 2.5, 5.0, 10.0, 12.5$ and $20.0$ km/s \cite{duchaineau-10}. Those results show the same general trends as the HYDRA simulations and the LEOS table for copper at a density of 15\% $\rho_{\rm bulk}$. Results for the molecular dynamics simulations of the foam ($p_c$, $\rho_c$, $T_c$) are also shown in figures~\ref{density-20kmps-pro}, \ref{pressure-20kmps-pro}, \ref{cufoam-hugoniot-pro} and \ref{temperature-20kmps-pro}.

Also shown in Fig.~\ref{d3-vfvp-pro} and listed in Table~\ref{uf-up-res} are results with an ideal vapor (IV) EOS for copper at ($p_0$, $\rho_0$, $T_0$) used in the jump condition, equations~(\ref{mass}), (\ref{momentum}) and (\ref{energy}).  The ideal vapor EOS is
\begin{equation}
p = \frac{\rho}{A} T ,
\label{ivp}
\end{equation}
and
\begin{equation}
\varepsilon = \frac{3}{2} \frac{p}{\rho} .
\label{ive}
\end{equation}
Here $A$ is the atomic mass of copper and is taken to be:
\begin{equation}
A = 1.055\times10^{-22} {\rm g} . 
\label{at-num}
\end{equation}

The results for $U_f$ using an ideal vapor EOS shown in Fig.~\ref{d3-vfvp-pro} and listed in Table~\ref{uf-up-res} were determined by solving equations~(\ref{mass}), (\ref{momentum}), (\ref{energy}), (\ref{ivp}), (\ref{ive}) and (\ref{at-num}) using the initial conditions of the foam and the values of $U_p$.  Results for the final state of the foam ($p_c$, $\rho_c$, $T_c$) as determined from the ideal vapor EOS are shown in figures~\ref{density-20kmps-pro}, \ref{pressure-20kmps-pro}, \ref{cufoam-hugoniot-pro} and \ref{temperature-20kmps-pro}.

Finally, shown in Fig.~\ref{d3-vfvp-pro} and listed in Table~\ref{uf-up-res} are results with a Van der Waals (VdW) EOS for copper at ($p_0$, $\rho_0$, $T_0$) used in the jump condition, equations~(\ref{mass}), (\ref{momentum}) and (\ref{energy}).  In reduced form, the Van der Waals EOS is
\begin{equation}
p = p_{\ast} \left[ \frac{\frac{8}{3}\frac{T}{T_{\ast}}}{\frac{\rho_{\ast}}{\rho}-\frac{1}{3}} - 3 \left( \frac{\rho}{\rho_{\ast}} \right)^2 \right] ,
\label{VdWp}
\end{equation}
and
\begin{equation}
\varepsilon = \frac{3}{2} \frac{T}{A} - \frac{1}{64} \frac{T_{\ast}^2}{p_{\ast}} \frac{\rho}{A^2} 1.60210000013 \times 10^{-22}.
\label{VdWe}
\end{equation}
Here the critical point of copper is taken from the LEOS tables to be:
\begin{eqnarray}
T_{\ast} = 5.27308 \times 10^{-4}~{\rm keV}, \nonumber \\
\rho_{\ast} = 1.98794~{\rm g/cm}^3 ,  \nonumber \\
p_{\ast} = 5.924738 \times 10^{-3}~{\rm Mbar} .
\label{critpnt}
\end{eqnarray}
The radius of a copper atom according the the Van der Waals equation of state and the values of the critical point is $r_{\rm Cu}^{\rm VdW} = 1.62 \times 10^{-8}$ cm.

The results for $U_f$ using the Van der Waals EOS shown in Fig.~\ref{d3-vfvp-pro} and listed in Table~\ref{uf-up-res} were determined by solving equations~(\ref{mass}), (\ref{momentum}), (\ref{energy}), (\ref{at-num}), (\ref{VdWp}), (\ref{VdWe}) and (\ref{critpnt}) using the initial conditions of the foam and the values of $U_p$.  Results for the final state of the foam ($p_c$, $\rho_c$, $T_c$) are also shown in figures~\ref{density-20kmps-pro}, \ref{pressure-20kmps-pro}, \ref{cufoam-hugoniot-pro} and \ref{temperature-20kmps-pro}.

The agreement between the ideal vapor, Van der Waals approximations, the HYDRA and molecular dynamics simulations and LEOS tables indicates that any reasonable estimate of the EOS will lead to a similar $U_p \left( U_f \right)$ dependence.

Figure~\ref{d3-vfvp-pro} and Table~\ref{exp-res} also show experimental results \cite{page-09}.  The initial densities of the foam in the experiments are not a perfect match for the foams simulated by HYDRA and molecular dynamics. However, the experimental results show similar $U_p \left( U_f \right)$ behavior to that observed in the LEOS tables ($\sim8 \%$ difference) and even the ideal vapor approximations ($\sim4 \%$ difference).  The Van der Waals approximation shows $U_p \left( U_f \right)$ behavior that are up to $\sim30 \%$ different from the experimental results.

\begin{table}[h]
\begin{center}
\begin{tabular}{|c|c|c|c|c|c|}
\hline
	$U_p$	&	$\rho_0$		&	$U_f$	&	$U_f$	&	$U_f$	&	$U_f$	\\
	(km/s)	&	(g/cm$^3$)	&	(km/s)	&	(km/s)	&	(km/s)	&	(km/s)	\\
			&				&Experiment	&	LEOS	&	IV		&	VdW		\\
\hline
	22.55	&	1.35			&	32.24	&	28.56	&	30.07	&	38.75	\\
	23.00	&	1.25			&	31.81	&	28.99	&	30.67	&	38.69	\\
	28.33	&	1.35			&	38.61	&	35.51	&	37.78	&	48.70	\\
	29.09	&	1.25			&	37.37	&	36.30	&	38.79	&	48.95	\\
\hline
\end{tabular}
\end{center}
\caption{Experimental and theoretical results for the front velocity $U_f$ as a function of piston velocity $U_p$.}
\label{exp-res}
\end{table}

\subsection{Density}

\begin{figure}[ht]
  \includegraphics[width=8.7cm]{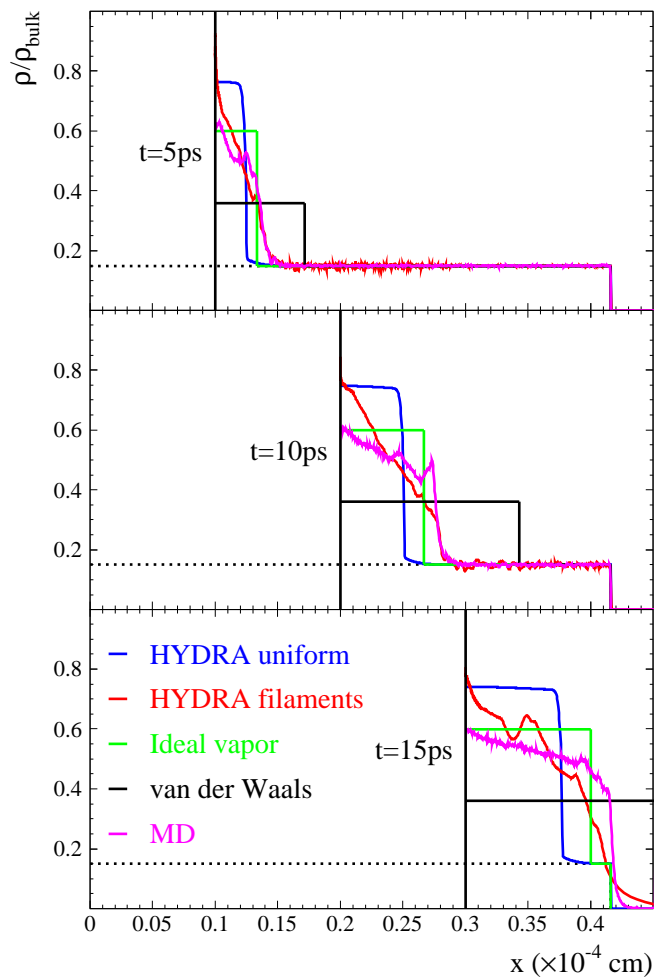}
  \caption{The density profiles at three different times are shown for simulations with $U_p = 20$ km/s.  The vertical black line shows the position of the piston.  The blue lines show the results of the simulations with the uniform foam. The red lines show the results of the simulations with foam made of filaments. The green lines show the results for the ideal vapor approximation.  The black lines shows the results for the Van der Waals approximation.  The magenta lines show the results of the molecular dynamics simulations. The dotted black line shows the density of the foam at time $t = 0$.}
\label{density-20kmps-pro}
\end{figure}

In the simulations as the piston moved from left to right it compressed the foam.  In the case of the HYDRA simulations of the uniform foam a shock wave was launched in the foam.  The foam was compressed to a density of $\rho_c = 73.9\pm0.1$\% $\rho_{\rm bulk}$ in the compressed region at break out time. Here the compressed region is defined as the region of the foam between the piston and the shock front. Figure~\ref{density-20kmps-pro} shows that that level of compression occurs shortly after the compression of the foam begins and a constant level of compression remains until the shock front Òbreaks outÓ into the vacuum region (or exits the foam region).

The results for the filament foam simulations were different.  Nearest to the piston the filament foam is compressed to a somewhat higher density than the compressed uniform foam simulations.  Away from the piston the density of the compressed foam drops with nearly constant slope to the density of the uncompressed foam. The plateau in density observed in the uniform foam simulation was never observed in the filament foam simulation.

The leading edge of the region of compressed foam, hereafter referred to as the compaction front, moved faster than the shock front of the uniform foam simulation which caused an earlier break out time for the filament foam simulation and thus a faster front velocity $U_f$.  The average density of the compacted region at break out time was $\left< \rho_c \right> \approx 70\pm10$\% $\rho_{\rm bulk}$. The average density of the compressed foam $\left< \rho_c \right>$ was obtained by dividing the total mass of copper between the piston and the end of the compressed region at break out time by the volume defined by those boundaries.  Here the compressed region is defined as the region of the foam between the piston and the compaction front. The error quoted reflects the RMS of the variation in the density in the zones of the compressed foam region.

The density of the compressed foam predicted by the ideal vapor EOS is 5.36 g/cm$^3$ which is lower than the HYDRA simulations and, due to the conservation of mass, leads to a faster front velocity.

The density of the compressed foam predicted by the Van der Waals EOS is 3.22 g/cm$^3$ which is lower than both the HYDRA simulations the ideal vapor approximation and, due to the conservation of mass, leads to a faster front velocity.

The average density of the foam in the compressed region in the molecular dynamics simulation shows results that are similar to the HYDRA simulations of the filament foam although the $\rho \left( x \right)$ slope for the MD simulation is less than the HYDRA simulation.

\subsection{Pressure}

\begin{figure}[ht]
  \includegraphics[width=8.7cm]{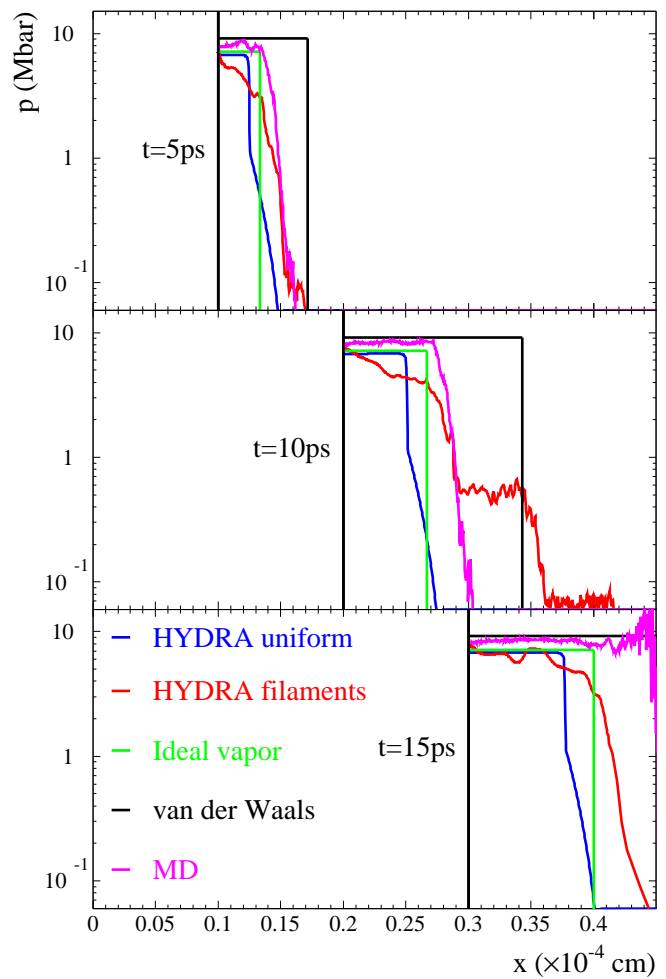}
  \caption{The pressure profiles at three different times are shown for simulations with $U_p$ = 20 km/s.  The colors and lines are the same as in Fig.~\ref{density-20kmps-pro}.}
\label{pressure-20kmps-pro}
\end{figure}

The HYDRA simulation of the uniform foam reached a mass weighted average pressure of $p_c = 6.77\pm0.02$ Mbar in the compressed region. Figure~\ref{pressure-20kmps-pro} shows that for the uniform foam simulation the pressure was reached shortly after the compression of the foam began and a constant level of compression remained until break out time.

The HYDRA simulation of the filament foam reached a mass weighted average value of $\left< p_c \right> = 7.1\pm0.8$ Mbar at break out time. Very close to the piston the pressure exceeded that attained by the uniform foam simulation.

The pressure of the compressed foam is 7.15 Mbar for the ideal vapor EOS,  9.19 Mbar for the Van der Waals EOS and $\sim$7.1 Mbar for the molecular dynamics simulation.  The molecular dynamics simulations (as well as the HYDRA simulations of the uniform foam) exhibits a nearly constant pressure throughout the compacted region.

\subsection{$p-\rho$ Hugoniot}

\begin{figure}[ht]
  \includegraphics[width=8.7cm]{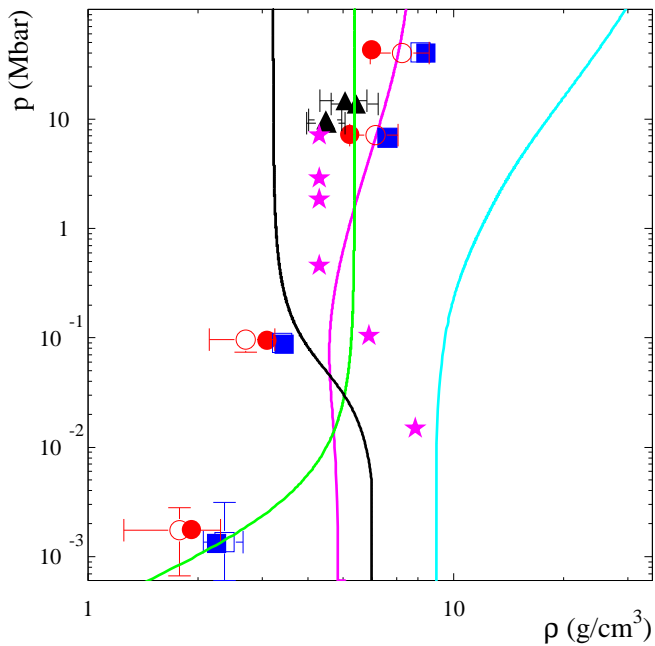}
  \caption{Shown above is the pressure as a function of density of the compressed region.  The symbols and colors are the same as in Fig.~\ref{d3-vfvp-pro}.  Empty symbols show direct measurements of the pressure and density while full symbols show estimates of the pressure and density using the jump conditions and values of the piston and front velocities.}
\label{cufoam-hugoniot-pro}
\end{figure}

The measured values of the density and pressure in the compressed region can be plotted in pairs and compared to the expected Hugoniot.  This is shown in Fig.~\ref{cufoam-hugoniot-pro}.

In general, the same results observed in Fig.~\ref{d3-vfvp-pro} are also observed in Fig.~\ref{cufoam-hugoniot-pro}.  However, plotting the data in this manner highlights the disagreement that was not easily observed in Fig.~\ref{d3-vfvp-pro}.

Figure~\ref{cufoam-hugoniot-pro} also shows two methods of determining the pressure and density for the HYDRA simulations.  In one method (shown with empty symbols in Fig.~\ref{cufoam-hugoniot-pro}) the average values of the pressure and density in the compressed region were measured directly from the density and pressure profiles of Figures~\ref{density-20kmps-pro} and \ref{pressure-20kmps-pro}.  In the other method (shown with full symbols in Fig.~\ref{cufoam-hugoniot-pro}) the density and pressure values were determined using the jump conditions and values of the piston and front velocities ($U_p$ and $U_f$).  The agreement between these two estimates (within error bars) confirms the conservation of mass and momentum.

Generally Fig.~\ref{cufoam-hugoniot-pro} shows that the two HYDRA simulations agree with each other and are close to the LEOS tables for copper at a density of 15\% $\rho_{\rm bulk}$ for pressures above 0.1 Mbar and are close to the experimental results. At the lowest piston velocities the agreement between the HYDRA simulation of the uniform foam and the LEOS tables is poor. The disagreement ($\rho_{\rm simulation} > \rho_{\rm table}$ and $p_{\rm simulation} > p_{table}$), at the lowest piston velocities, between the LEOS tables and the HYDRA simulations of the filament foam may arise due to the vapor precursor taking away some of the mass of the copper, thus lowering the density. The vapor precursor also increased the temperature of the filaments (see Fig.~\ref{temperature-20kmps-pro}) and increased the pressure.

The Van der Waals approximation and the molecular dynamics simulation show very similar behavior with an offset in the density.  This offset arises from the differences in sizes of the atoms in each system.  As stated above, the radius of a copper atom in the Van der Waals approximation is $1.62 \times 10^{-8}$ cm while in the molecular dynamics simulation it is $1.45 \times 10^{-8}$ cm.  The larger atoms in the Van der Waals approximation lead to a lower {\it limiting} value of the density in the low pressure ($p \lesssim 10^{-2}$ Mbar) region than for the molecular dynamics simulation. For the high pressure region ($p \gtrsim 1$ Mbar) both the Van der Waals and molecular dynamics simulation tend to constant values of density just as the ideal vapor.  In that respect, in the high pressure and lower pressure region, both the Van der Waals and molecular dynamics simulation exhibit ideal behavior.  The HYDRA simulations and LEOS tables do not exhibit this sort of behavior over the range of pressures and densities examined in this work.

\subsection{Temperature and ionization}

\begin{figure}[ht]
  \includegraphics[width=8.7cm]{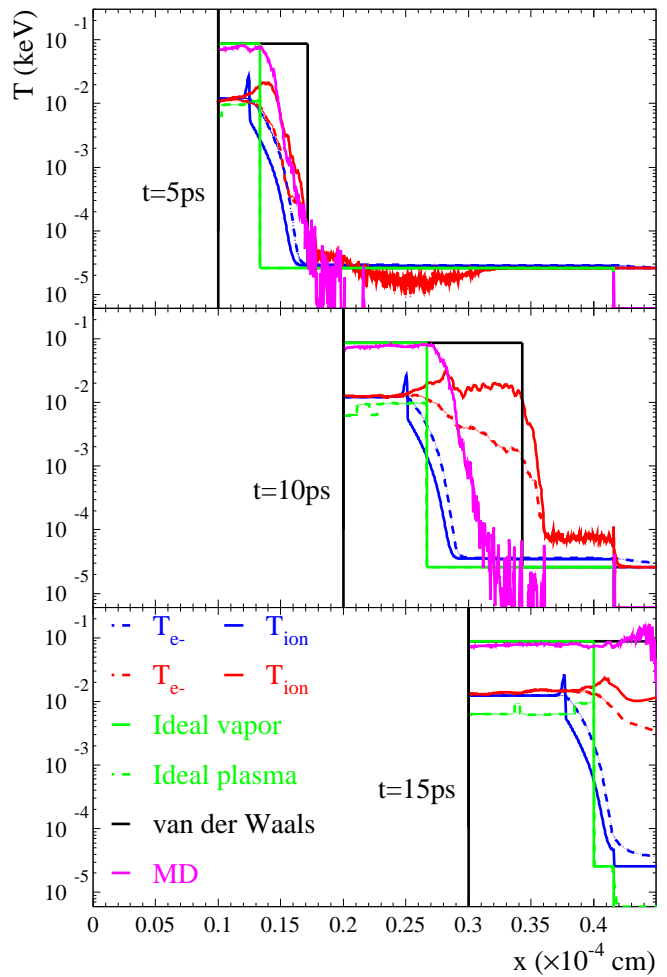}
  \caption{The temperature profiles at three different times are shown for simulations with $U_p = 20$ km/s.  The colors and lines are the same as in Fig.~\ref{density-20kmps-pro}.  Results for the ion and electron temperatures from the HYDRA simulations and estimates for an ideal plasma are shown.  See text for details.}
\label{temperature-20kmps-pro}
\end{figure}

In all the HYDRA uniform simulations in the compressed region the ions were heated to $0.01234\pm0.00004$ keV and the electrons to of $0.01235\pm0.00005$ keV.  In the HYDRA filament foam, the average values of the temperatures in the compressed region were $0.0134\pm0.0007$ keV for the ions and $0.0133\pm0.0005$ keV for the electrons.  All temperatures given above are mass weighted average results.

The values of the temperature for the ideal vapor approximation (0.08875 keV), the Van der Waals approximation (0.08771 keV) and the molecular dynamics simulations (0.077$\pm$0.003 keV, the average and RMS of the temperature in the compacted region at time $t = 15$ ns) are all similar and nearly an order of magnitude larger than the HYDRA results.  It is shown below that this is due to the presence of ionization in HYDRA and the absence of ionization in the other systems.

An ideal approximation can be made for a plasma just as for a vapor \cite{drake-06}.  To go from an ideal vapor to an ideal plasma, the average number of electrons ionized from an atom of the plasma, $\left< Z \right>$, is the key.  The pressure of the ideal plasma depends on the pressure and temperature as follows
\begin{equation}
p_{\rm P} = \frac{\rho_{\rm P} \left( 1 + \left< Z \right> \right)}{A} T_{\rm P}
\label{pion}
\end{equation}
Here the subscript ``P'' denotes quantities for the ideal plasma.  The energy of the ideal plasma is then given as
\begin{equation}
E = \frac{3}{2} N \left( 1 + \left< Z \right> \right) T_{\rm P} + N e_i
\label{eion}
\end{equation}
$N$ is the number of atoms in the plasma and $e_i$ is the energy required to ionize $\left< Z \right>$ electrons from an atom.  Values of $e_i$ were determined from the enthalpies associated with ionization for copper given in Table ~\ref{cu-enth} \cite{webelm}.

\begin{table}[h]
\begin{center}
\begin{tabular}{|r|r|}
\hline
	Ionization	&	Enthalpy		\\
	energy	&	(kJ/mol)		\\
	number	&				\\
\hline
	1	&	745.5	\\
	2	&	1,957.9	\\
	3	&	3,555.0	\\
	4	&	5,536.0	\\
	5	&	7,700.0	\\
	6	&	9,900.0	\\
	7	&	13,400.0	\\
	8	&	16,000.0	\\
	9	&	19,200.0	\\
	10	&	22,400.0	\\
	11	&	25,600.0	\\
	12	&	35,600.0	\\
	13	&	38,700.0	\\
	14	&	42,000.0	\\
	15	&	46,700.0	\\
	16	&	50,200.0	\\
	17	&	53,700.0	\\
	18	&	61,000.0	\\
	19	&	64,702.0	\\
	20	&	163,700.0	\\
	21	&	174,100.0	\\
\hline
\end{tabular}
\end{center}
\caption{Copper ionization enthalpy values.}
\label{cu-enth}
\end{table}

If the ideal vapor and the ideal plasma are at the same initial conditions and are compressed by pistons moving with the same velocities, conservation of energy allows us to combine equations~(\ref{ive}) and (\ref{eion}) and solve for $T_{\rm P}$, the temperature of the ideal plasma as
\begin{equation}
T_{\rm P} = \frac{T - \frac{2}{3}e_i}{1+ \left< Z \right>} .
\label{ion-temp}
\end{equation}
This shows the the temperature of the (ideal) plasma is always less than or equal to the temperature of the (ideal, unionized) vapor and agrees with the behavior shown in Fig.~\ref{temperature-20kmps-pro}.

\begin{figure}[ht]
  \includegraphics[width=8.7cm]{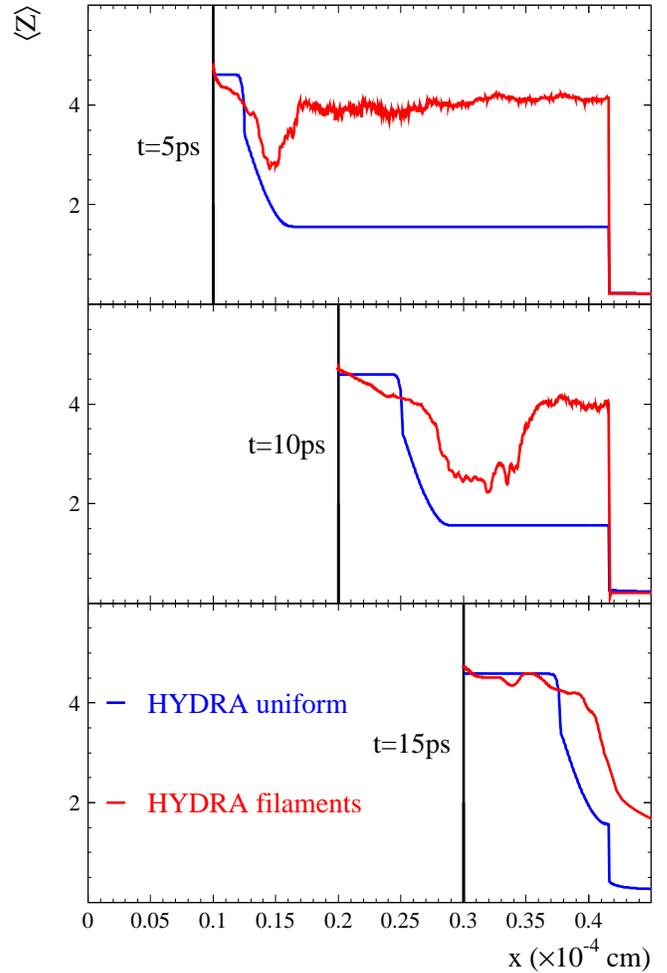}
  \caption{The ionization profiles at three different times are shown for simulations with $U_p$ = 20 km/s.    The colors and lines are the same as in Fig.~\ref{density-20kmps-pro}.}
\label{ionization-20kmps-pro}
\end{figure}

Using the HYDRA simulations estimates of $\left< Z \right>$ (shown in Fig.~\ref{ionization-20kmps-pro}) and the temperature values of the ideal vapor (shown in Fig. ~\ref{temperature-20kmps-pro}) gives the results for $T_{\rm P}$ shown in Fig.~\ref{temperature-20kmps-pro}.  This estimate of the temperature of the compressed foam agrees better with the HYDRA results than do the ideal vapor estimates demonstrating that the increased temperature observed in the ideal vapor approximation, the Van der Waals approximation and the molecular dynamics simulation arises from a lack of ionization.

\begin{figure*}[ht]
  \includegraphics[width=17.9cm]{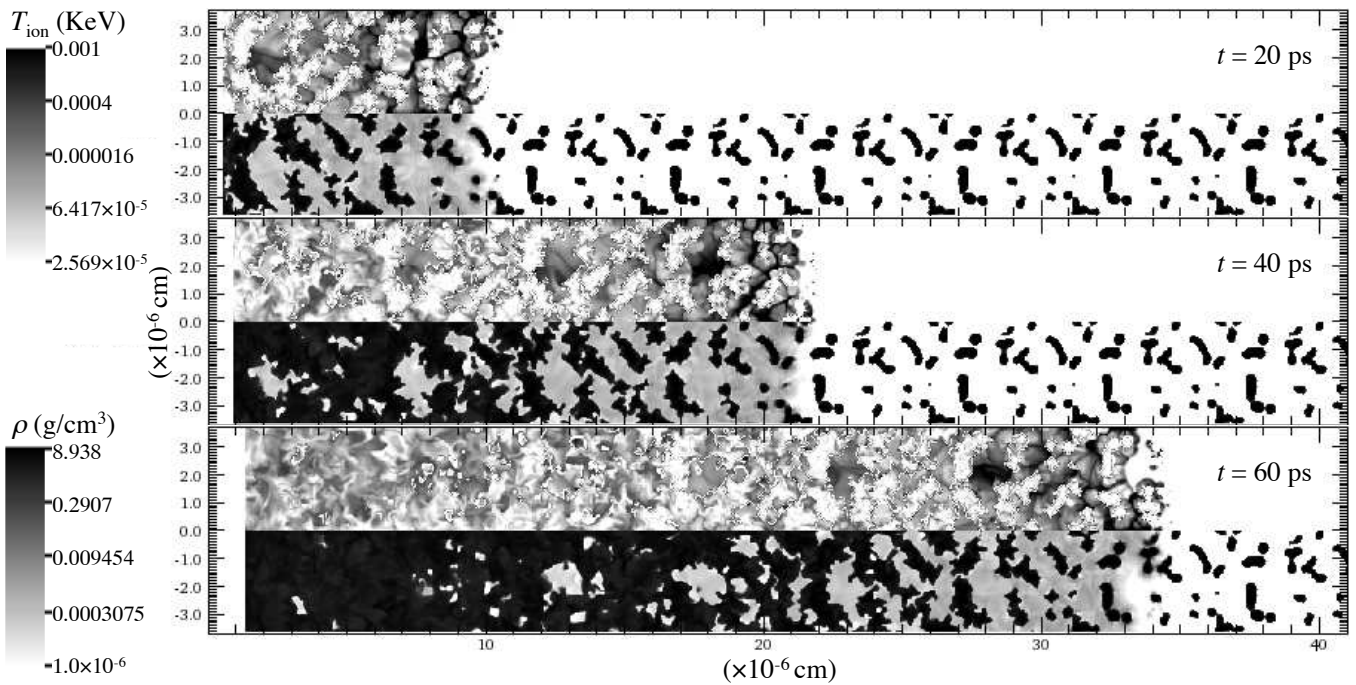}
  \caption{Snapshots of a slice through the filament foam simulation at times $t = 0$, 20, 40 and 60 ps. The top portion of each plot shows the ion temperature (keV) while the bottom shows the density (g/cm$^3$); see gray (log) scales at the left. The foam is being compressed by a piston moving at 0.2 km/s in the $+x$-direction. The filaments are heated well beyond the compacted foam region by the vapor precursor. The speed of the vapor precursor is approximately 5.5 km/s, which is a mach number of about 1.4.}
\label{foam-melt-0}
\end{figure*}

\subsection{Ahead of the compaction front}

The previous sections described behavior in the compacted region of the foam. This section describes the behavior of the foam ahead of the compaction front.

The filament foam simulations showed regions of high pressure ahead of the compaction front.  This is evident in the middle panel of Fig.~\ref{pressure-20kmps-pro} which shows the simulation at a time of $t = 10$ ps.  At that time the compaction front had progressed to a position of nearly $x \approx 3\times10^{-5}$ cm, but a region of constant pressure of approximately 0.5 Mbar is observed between the end of the compaction front and about $x \approx 3.5\times10^{-5}$ cm.  This region of high pressure reflects the extent of the so-called vapor precursor. The vapor precursor arises because as the piston compressed the filament foam, the filaments were heated and some copper ablated from the surface of the filaments and formed a high temperature vapor.  The vapor then flowed through the voids in the foam between the filaments ahead of the compaction front with a velocity of approximately 36 km/s for the $U_p = $ 20 km/s simulation. The vapor precursor heated the filaments ahead of the compaction front and increased the average pressure of the foam. The uniform foam simulations show no vapor precursor since there are no voids for any vapor to travel through.

Beyond the leading edge of the vapor precursor region there is another region of constant pressure of about 0.07 Mbar.  The region extends from the end of the vapor precursor region to the end of the foam $3.5\times10^{-5} {\rm cm} \lesssim x \lesssim 4\times 10^{-5} {\rm cm}$ and will be called the preheated region.  The pressure is elevated in this region because the high temperature of the crushed filaments has been conducted through the filaments to the end of the foam.  All of the foam ahead of the compaction front was heated in this manner.  Examining the temperature profiles shown in Fig.~\ref{temperature-20kmps-pro} shows this behavior as well. The preheat front moves through the filament foam at a velocity of approximately 60 km/s for the simulations with $U_p = $ 20 km/s. This preheating behavior was observed in only the simulation with higher piston velocities.  At lower piston velocities, the filaments were not heated sufficiently to preheat the material ahead of the vapor precursor.

\begin{figure*}[ht]
  \includegraphics{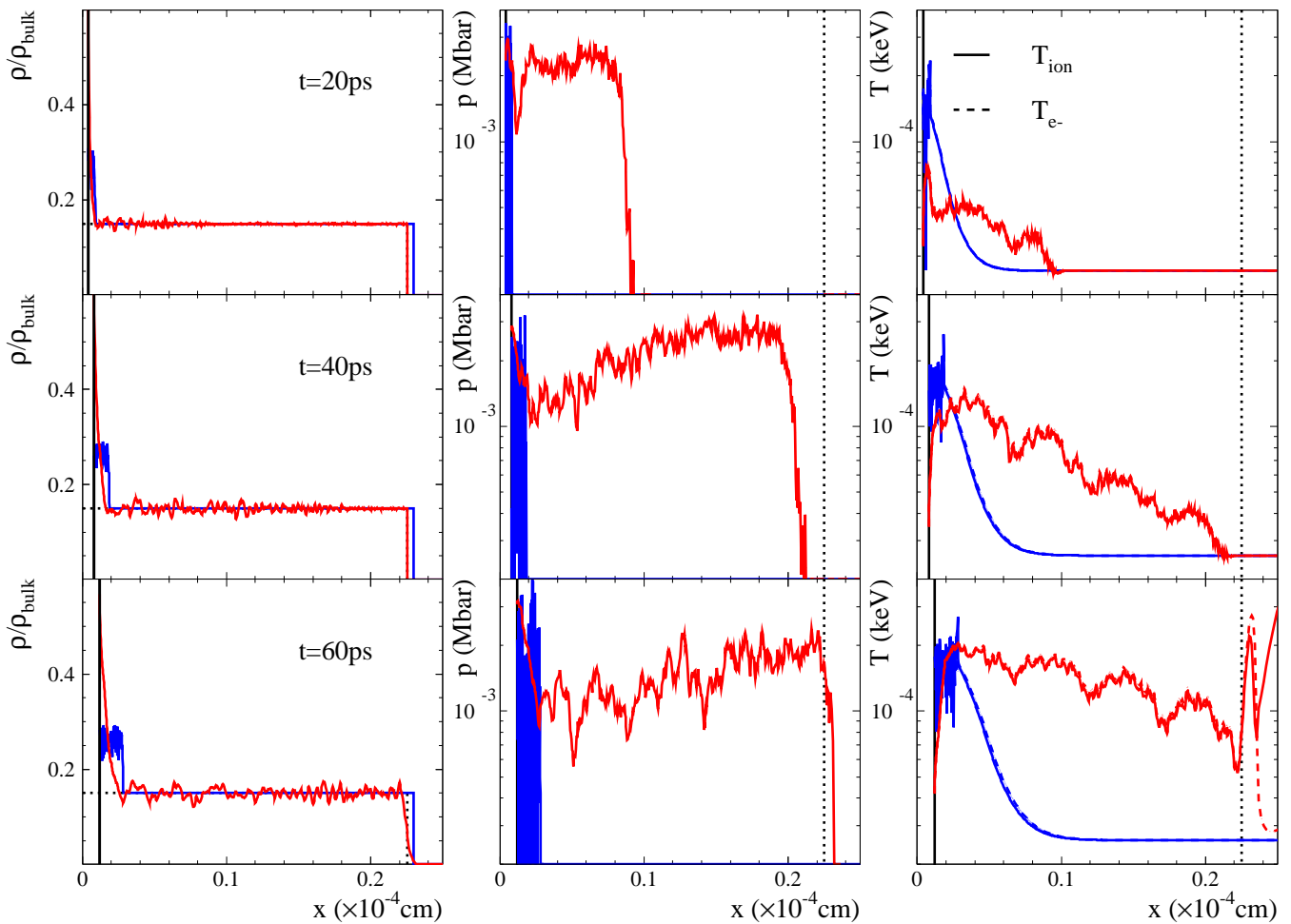}
  \caption{From left to right: the density (g/cm$^3$), pressure (Mbar) and temperature (keV) profiles at times $t = 0$, 20, 40 and 60 ps. The foam is being compressed by a piston moving at 0.2 km/s in the $+x$-direction. The length of the foam slab is $l_{\rm foam} \approx 23\times10^{-6}$ cm. The vapor precursor region is clearly present, however the preheated region is not. The colors and lines are the same as in Fig.~\ref{density-20kmps-pro}. The vertical dotted line shows the end of the foam slab at time $t= 0$.}
\label{vapor}
\end{figure*}

\begin{figure}[ht]
  \includegraphics[width=8.7cm]{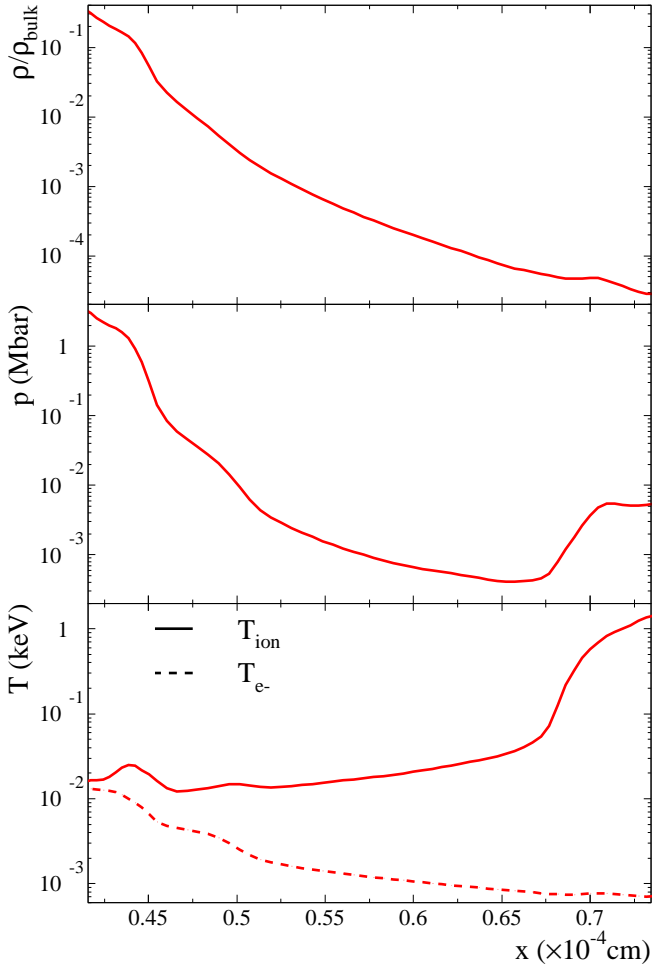}
  \caption{From top to bottom: the density, pressure and temperature profiles for the HYDRA filament simulation $U_p = 20$ km/s at $t = 16$ ps. The plots show $x$ starting at the $t=0$ position of the foam/void interface.  The rise in all quantities after $x = 0.7 \times 10^{-4}$ cm is due to the vapor hitting the end of the simulation at $x = 0.75 \times 10^{-4}$ cm and bouncing back.}
\label{vapor-pre-20kmps}
\end{figure}

\begin{figure}[ht]
  \includegraphics[width=8.7cm]{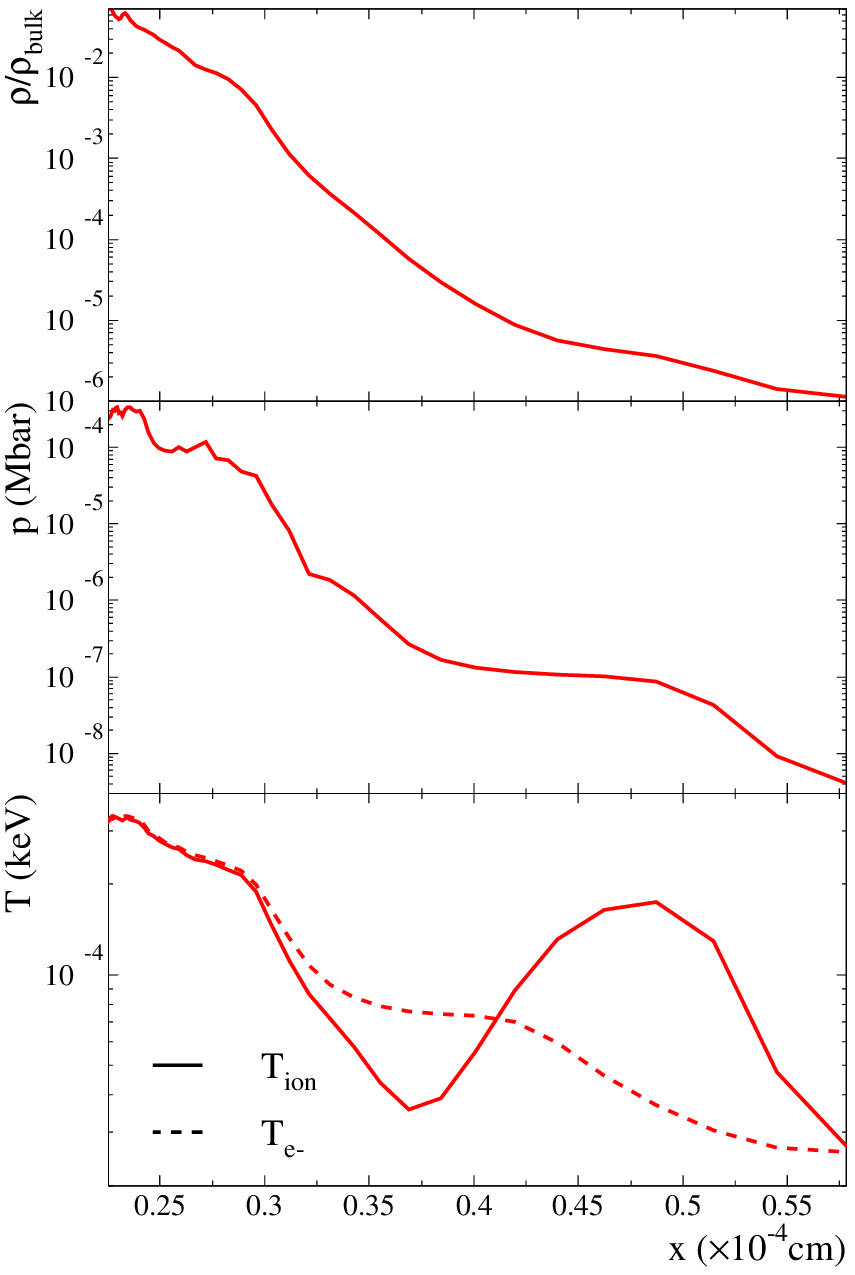}
  \caption{From top to bottom: the density, pressure and temperature profiles for the HYDRA filament simulation $U_p = 0.2$ km/s at $t = 260$ ps.  The plots show $x$ starting at the $t=0$ position of the foam/void interface.}
\label{vapor-pre-002kmps}
\end{figure}

However, it is not clear that the extent of this preheat is entirely physical for the filament foam.  The molecular dynamics simulations suggest that when heated the filaments melt and then the melted portion of the filaments and are severed from the solid filaments due to surface tension effects\cite{duchaineau-10}. This behavior arises due to the small radius of the filaments (and large surface to volume ratio). The melting and severing would preclude the heating ahead of the vapor precursor observed in the HYDRA simulations of the filament foam.  This may be expected since HYDRA and the LEOS tables ignore surface effects.

For simulations at slower piston velocities the features of the vapor precursor and the preheating are combined. Figure~\ref{foam-melt-0} shows snapshots of a slice through a filament foam simulation with a piston velocity of 0.2 km/s. Figure~\ref{vapor} shows the density, pressure and temperature profiles for a similar simulation. The snapshots and profiles show that the filaments are heated in regions well ahead of the compacted foam region (the compacted region for these simulation and times is $x \lesssim 3\times10^{-6}$ cm) by the vapor precursor. The vapor precursor moves through the foam at a velocity of about 5.5 km/s. The filaments then melt and flow together forming a liquid with the same average density as the foam. This liquid is then compressed by the piston.

Once the vapor precursor exits the foam region it acts as a freely expanding vapor. For a $U_p = 20$ km/s the front of the vapor precursor in vacuum moves at approximately 70 km/s. Figure~\ref{vapor-pre-20kmps} shows the density, pressure and temperature profiles for the vapor precursor in vacuum for $U_p = 20$ km/s. For a $U_p = 0.2$ km/s the front of the vapor precursor in vacuum moves at approximately 2 km/s. Figure~\ref{vapor-pre-002kmps} shows the density, pressure and temperature profiles for the vapor precursor in vacuum for $U_p = 2$ km/s.

\section{Conclusions}

The HYDRA simulations, LEOS tables, ideal vapor approximation, Van der Waals and molecular dynamics simulations well reproduce the $U_p \left( U_f \right)$ dependence observed in experiment and HYDRA simulations.  This indicates a lack of sensitivity in $U_p \left( U_f \right)$ to simulation method or equation of state for copper nano-foams. Similarly, equations of state and methods of simulation have little effect on the values of the pressure in the compacted region. The density in the compacted region is observed to be the same for the HYDRA simulation with filaments and the molecular dynamics simulations, but different for the HYDRA simulations of the uniform foam and the ideal vapor and Van der Waals approximations. Ionization is seen to have a large effect the compressed foam temperature which makes is suitable for testing equations of state. The filament foam simulations exhibit a vapor precursor that affects the foam and precedes the compaction front and is absent in the uniform foam simulations.

\section{Acknowledgments}

The authors would like to thank Farid Abraham for his input on the filament foam simulations.

This work performed under the auspices of the U.S. Department of Energy by Lawrence Livermore National Laboratory under Contract DE-AC52-07NA27344.

\end{document}